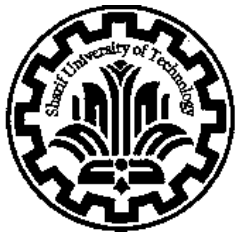

# Sharif University of Technology

# Nonlinear behavior of RC shear walls: From experiments to the field reports

**Mojtaba Harati**

University of Science and Culture

**Mohammadreza Mashayekhi**

Sharif University of Technology

**Ali Khansefid**

Khaje Nasir Toosi University of Technology

**Saeid Pourzeynali**

**Arash Bahar**

University of Guilan

# Nonlinear behavior of RC shear walls: From experiments to the field reports

**Mojtaba Harati**

University of Science and Culture

**Mohammadreza Mashayekhi**

Sharif University of Technology

**Ali Khansefid**

Khaje Nasir Toosi University of Technology

**Saeid Pourzeynali**
Arash Bahar

University of Guilan

# ABSTRACT

Post-earthquake damage evaluation has indicated that although the buildings with shear walls exhibited an appropriate overall performance in the recent sever earthquakes, in some cases, however, the columns and the shear walls were damaged, due to presence of short structural elements and inadequate transverse reinforcement. Such memories amongst engineers have promoted this attitude that the shear walls structures exhibit a brittle overall performance. So the conviction that shear walls are inherently brittle still prevails amongst engineers, and therefore, they usually prefer to choose moment resisting RC frames without shear walls. For getting an insight into the nonlinear behavior of RC shear walls, we have reviewed the experiments conducted on the RC shear walls. Besides, through using the post-earthquake reports written after major earthquakes, we have checked and scrutinized the seismic performance of RC frames equipped with shear walls too. Investigations performed on the nonlinear behavior of the shear walls have demonstrated that slender shear walls, designed to fail in flexural mode, could safely dissipate excitations energy of the earthquakes. These studies also revealed that even squat shear walls can be designed in a way that their behavior would be similar to those of the slender walls. Considering these matters, designed shear walls according to the recent codes maintain sufficient shear strength and respectable energy dissipation capability.

# ACKNOWLEDGMENTS

The authors would like to sincerely acknowledge Dr. Omid Sedehi, the Research Associate at the University of Hong Kong, for his sincere supports, cooperation, motivations and advices. The authors also acknowledge valuable discussions on the research with Dr. Afshin Mohammadi, Dr. Sayed Ali Mirfarhadi, Dr. Homayoun Estekanchi and Dr. Hassan Moghaddam. At the end, the authors are particularly proud to dedicate this research to the people who lost their families and relatives in Manjil-Rudbar earthquake of 1990.

# CONTENTS

# LIST OF FIGURES

# 1- Introduction

Shear walls in RC structures are usually used as a means to withstand against the lateral forces of an earthquake. Due to their large stiffness and strength compared to concrete frames, these members absorb a considerable amount of the base shear, and it seems that the name "shear wall" for these structural members is not due to shear behavior (these elements in walls with relatively large height—used in modern structures—mainly act as a bending element), but also because of the considerable amount of base shear force they can readily absorb. For up to 20-storey structures, this member is a matter of choice for structural designers of the buildings. But for structures with more than 30 floors, the designer will have to use these members for economic reasons and lateral drift control [1]. The use of shear walls reduces the earthquake-induced deformation of the entire structure as well as the deformations and forces of structural members [24, 25], including the deformation of the shear wall itself [2, 26].

There are codes and views that still hold this opinion that RC shear walls behave in a brittle manner. For this reason, a number of existing codes suggest a lower ductility factor for designing the shear wall structural systems than that for frame systems [1, 2]. In this case, the nonlinear behavior of shear walls would be examined in this technical report. For this purpose, experiments on shear walls over the past 20 years will be used, where the main source of the experiments are papers published in the ASCE and ACI structural journals as well as the latest World Conference on Earthquake Engineering.

In this technical report, we will first talk about the general categories as well as the design procedures of the tested shear wall specimens. Then our attention will be directed towards laboratory equipment and different types of loading regimes. Subsequent discussions will focus on the types of failure mechanisms and cracks in RC shear walls. The general definition of failures observed in the walls is also discussed. Next, we will talk about the shear walls and their acceptance criteria in the performance-based design codes. Finally, the seismic performance of the shear wall structures is reviewed from the post-earthquake field reports.

## 2- General characteristics of shear walls

The shear wall is a member that is made of reinforced concrete, which is designed to withstand earthquake lateral loads. These structural members have horizontal (shear), vertical (flexural), and inclined reinforcements (to resist and reduce shear slip between the foundation and the wall). According to Figure 1, they are made of rectangular sections and designed or constructed in two forms—with and without boundary elements. Vertical reinforcing net(s) is placed individually or in two opposite faces. In some cases, vertical reinforced meshes with welded joints are also used.

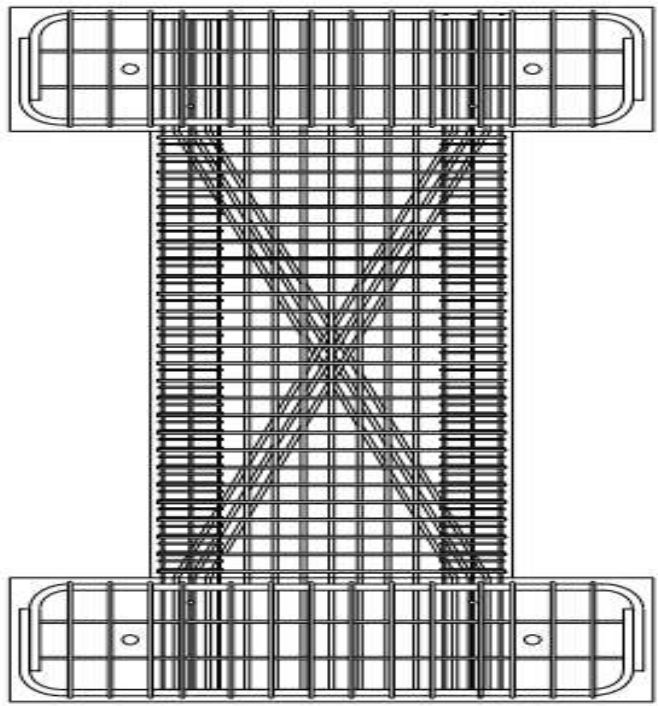

Figure 1. Arrangement of the steel reinforcements in an RC shear wall [24]

For constructing and designing shear wall specimens, researchers choose different parameters independently to interpret the shear wall test results:

1- The amount and percentage of longitudinal reinforcement and the distance between them.
2- The amount and percentage of diagonal reinforcements and the level they are placed along the wall height.

3- Examining the presence or absence of boundary elements in shear walls.
4- Changing or adjusting applied vertical loads of the walls; this value is usually equal to a percentage of the total cross-sectional area of the shear wall:

$$\lambda_p = \beta fc\, Ag \qquad (1)$$

where coefficient β is less than one

5- Altering the dimension ratio of the walls; in RC shear walls, the ratio of height to length is called the wall dimension ratio:

$$\alpha_r = h/B \qquad (2)$$

In the above formula, h is the height of the wall and B is the longitudinal dimension (on the plan) of the rectangular cross-section of the wall. This parameter affects the lateral deformation behavior of the shear wall. For a dimension ratio of less than 1, the shear wall deformation pattern is mainly in shear form though shear wall failure may, in any case, be based on the bending or flexural strength. For dimension ratios greater than 2, shear wall performance is usually predominately associated with the bending behavior (Figure 2) [10, 12]. According to our literature review, it seems that walls with a dimension of 1.5 are also classified as the shear walls with a shear manner.

6- Shear span ratio; in shear walls, the ratio of load height to length of the wall is called the shear span ratio:

$$\alpha_s = h'/B \qquad (3)$$

In the above formula, h' is the height of the lateral load P; and B is the longitudinal dimension of the rectangular cross-section of the shear walls on the plan. This parameter affects the lateral deformation behavior of the shear wall. This coefficient would be equal to the dimension ratio of the walls if the lateral load P is applied at the top of the shear wall (Figure 2) 7- Materials are changed in specimens of the walls ($f'_c, f_y$); usually the properties of the materials used in shear walls are obtained directly from the experiment

8- Changing the factor stands for shear-compression ratio; this load ratio is defined as follows [4]:

$$\lambda_p = \frac{V_{max}}{fc.Ag} \tag{4}$$

where $V_{max}$ is the maximum shear force applying to the wall

9- Construction joint at the base of the shear wall and investigation of its Impact on shear sliding deformation.

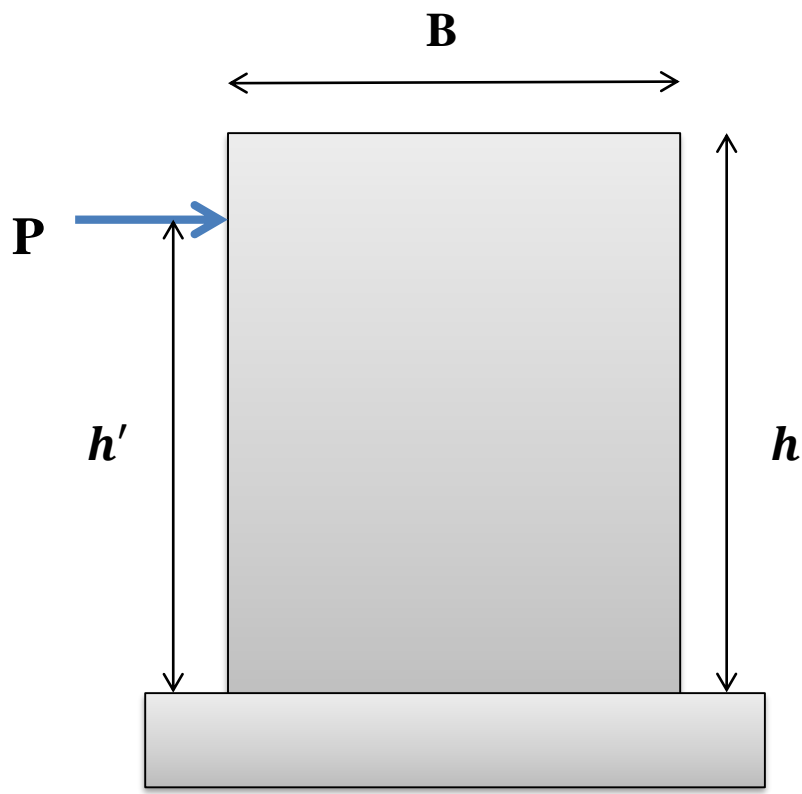

Figure 2. Parameters used in defining dimension and shear-span ratios [24]

# 3- Design of shear walls

In shear walls, there are generally two types of structural behavior, the bending and the shear behaviors. In RC structures and for the design of shear walls, members are usually designed to exhibit a flexural behavior, where the wall member is designed to provide energy absorption and energy dissipation by forming a plastic hinge at the foot of the wall. To prevent undesirable modes of failure, building codes set their criteria for the formation of a full plastic hinge [3]. However, the reference [3]—quoting the words by Lefas et al. [7], Tremblay et al. [8], Panneton et al. [9]—reported that the plastic hinges can also emerge at the top of the wall.

The nonlinear dynamic behavior of shear walls is controlled by several factors that can be quite vague and complex. These contributing factors can include the effects of superposition of dynamic vibrational modes in non-elastic ranges, post-yielding behavior of shear walls under increasing dynamic loads and the effect of bending-shear-axial demand interaction as well as the random effect of the acceleration of the ground motions. In general, the effect of the source of seismic excitations can make a great deal of difference to the assumptions made earlier. Only a limited numbers of building codes have considered this matter [3].

The Figure 3 illustrates the cyclic curve of displacement versus shear force for the top of the wall level. As shown in the Figure 3, the area under the curve is greater for bending behavior (the figure at the right side, Figure 3 (b)) than for shear behavior, indicating greater absorption and energy loss gains for the member with bending behavior. One of the prominent features of members with shear behavior can be seen in their cyclic curves that are associated with the pinching occurring at the center of the graph. This phenomenon has been reported as a sign of a problem originated from a lack of energy dissipation mechanism and capability in the closure process of shear cracks. It should be noted that shear-bending behavior can also be seen in the shear wall.

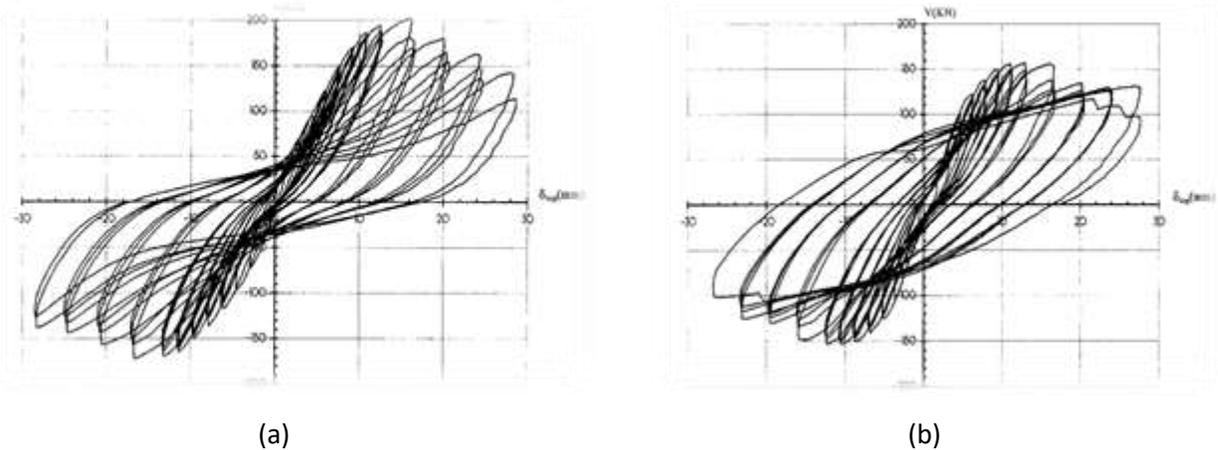

(a)                          (b)

Figure 3. Shear cyclic behavior and the bending cyclic manner of the RC walls [11]

The shear walls are able to reach their ultimate strength in both shear and bending modes. Even when the force-displacement behavior of a member is dominated with shear behavior (top left), the member eventually reaches its flexural strength [11] in case the flexural strength of the member is greater than its shear strength. It is worth mentioning that the cyclic deformation behavior of the shear wall is not only a function of the bending or shear strength of the wall cross-section and depends on several other factors.

Wall specimens tested for research works and scientific papers are designed and prepared for the experimental test in the following ways:

1- They were designed using valid building codes and their reinforcement details are carefully configured in accordance with the rules mentioned in the relevant codes.

2- The shear walls were designed using displacement-based codes or the rules from the performance-based design method.

3- In terms of the amount and detail of steel reinforcement, researchers design and construct test samples that are very similar to the walls that are locally constructed.

# 4- Laboratory equipment

Laboratory equipment for shear wall experiments is divided into two categories:

## 4-1- Stationary equipment

In this category, there are stationary laboratory equipment. These equipments are as follows:

### 4-1-1- Strong floor

This floor has high strength, and researchers use this floor to fasten the rigid beam of the shear wall to the base. This rigid beam is used as the foundation of the test specimens of the RC walls. The reinforcement details of these members (rigid beams beneath the wall) are chosen in a way that they act as a rigid element.

### 4-1-2- Strong wall

This wall has high strength, which is used as a support for lateral load apparatus (e.g. hydraulic jack) that is mounted to the top rigid beam of the shear wall. The reinforcement details of this top beam are similar to those beams considered beneath the RC wall. It should be noted that this beam can be made of steel in some cases too.

### 4-1-3- Vertical and horizontal load devices

These devices are used to apply vertical and horizontal load to the shear wall. These devices work in both force and displacement control modes. The loading capacity of these load-transfer devices is different from each other.

### 4-1-4- Frame incorporating vertical and horizontal loading devices

These steel frames actually have a similar function to a strong wall and serve as a support for horizontal and vertical jacks. Steel frames are also used in cases where multi-story shear walls are tested. Unlike their use in squat shear walls as a support for hydraulic jacks, the mentioned steel frames are commonly utilized as lateral retaining structures in multi-story RC walls. These structures are expected to perform two basic tasks. First, they prevent out-of-plane deformations of the shear wall. And secondly, they can transfer seismic mass to the RC wall. Concerning the seismic mass transfer performance of these frames, you can refer to a section of this report prepared for the dynamic loading of shear walls (Figure 5 and Figure 10).

**4-1-5- Shake tables**

They are used to model earthquake excitations. More details about these tables are provided in a section that is provided for the dynamic loading (Figure 10).

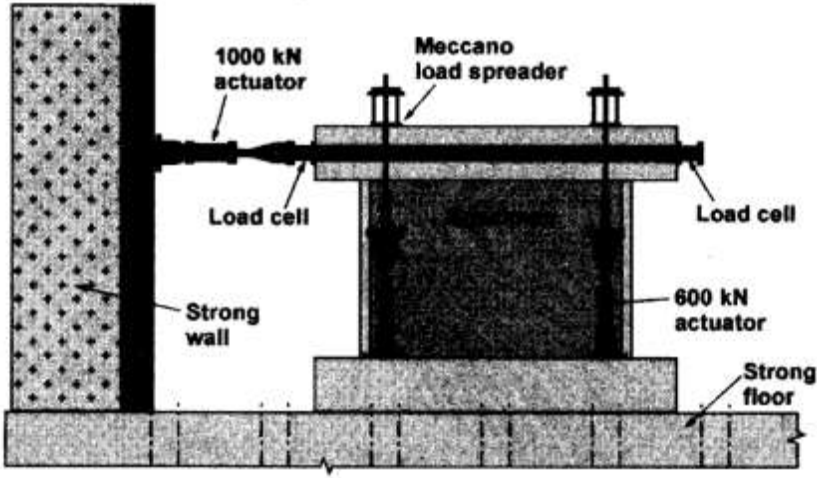

Figure 4. Laboratory stationary equipment, strong wall [5]

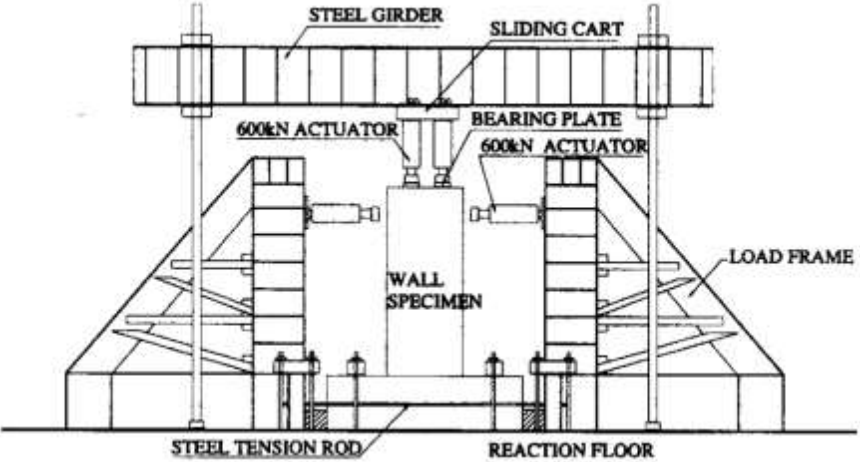

Figure 5. Laboratory equipment, the steel frame [5]

It should be noted that RC walls normally designed and constructed based on the size and dimensional limitations of stationary laboratory equipment. So, researchers obtain the dimensions and scales of the samples according to the available equipment in the laboratory environment.

**4-2- Non-stationary or portable equipment**

These types of equipment are usually placed on the shear walls, and they are used to record and control response characteristics such as the strain of reinforcements, concrete strain as well as structural displacements.

**4-2-1- Tools to read the load level**

These devices (Load Cells, Pressure Transducer) are used to read the load level. These devices are also used when loading protocol is intended to be force-control. These devices track and record the load or displacement and measure the amount of forces transmitted between two points these tools are set to work.

**4-2-2- Devices to read displacements**

For various reasons such as calculation of maximum wall displacement, measuring the curvature, the extent to which base of the walls slips, finding shear displacements, devices to read structural displacements are used to record displacement histories of different points of the shear wall during experimental testing. Various devices may be employed to read such displacements. X-shape strain gauges or displacement meters are used to calculate shear wall deformations (Figure 6). A sliding gauge is used horizontally to record the slips occurring at the bases of the RC wall (in some cases it can be used to record relative slip between the wall and foundation, especially when there is a construction joints in place). To calculate the displacement on the top edges of the RC wall (for curvature calculation) and out-of-plane deformations, displacement meters are taken to be at work in the appropriate locations and in the suitable directions.

**4-2-3- The ways to record strains**

Strain gauges are also used to read strains in concrete and shear wall reinforcement.

### 4-2-4- Measuring wall cracks

Photography and video cameras are used to record wall cracks [6]. When the camera is used for this purpose, the test at each round of loading is paused and the propagation of the witnessed cracks is recorded.

### 4-2-5- Measurement of the floor acceleration

Acceleration meters are used to record floor acceleration which is typically recorded to be utilized in other research studies. For example, to plot the acceleration distribution against the floor levels or for the analyses that require computational verification on the acceleration of each story level.

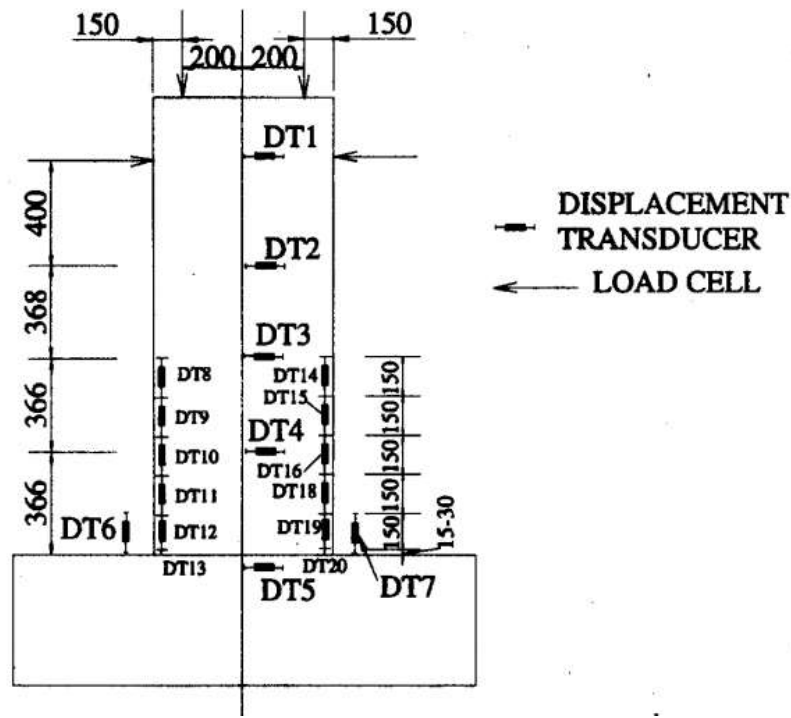

Figure 6. Portable equipment mounted on the RC wall [7]

In the figure above, as seen, the strain gauges are concentrated at the foot of the wall. This is mainly due to the importance of strains that may emerge around the area of the plastic hinges.

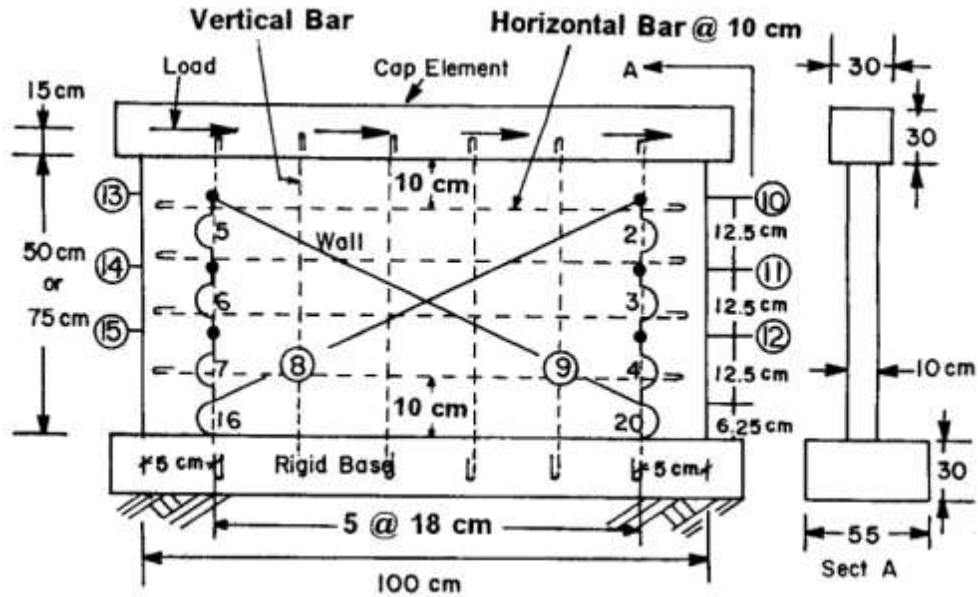

Figure 7. Portable wall-mounted strain gauges 8 and 9, which are utilized to read and to record shear deformations [10]

## 5- Loading regimes

The shear walls are subjected to the following loads in the laboratory environment:

**5-1- Vertical loading**

Vertical loading is usually equal to a percentage of wall strength that is calculated using the total cross-sectional area of the shear wall.

$$\lambda_p = \beta fc\, Ag \tag{5}$$

where coefficient β is less than one theoretically, which had different values in the following experiments:

Previous works published over the last 30 years indicate that the β ratio of the applied vertical load to RC walls was chosen to be less than 0.15; there were only three walls in the literature with a vertical load ratio (coefficient β) of more than 0.4 [4]. However, it was reported that the value of this ratio for the RC walls used in tall structures constructed in China is between 0.3 and 0.6. While building codes do not directly report an upper band for this ratio [4], the upper limit of this parameter in different codes seems to be less than 0.4 according to observations and reviews of the authors of this technical report. This type of loading is usually done in an instant way and are controlled by force.

**5-2-    Lateral loading**

For lateral loading, forces are usually applied along the wall plane and are put around the strong axis of the shear wall. Lateral loading on the RC shear walls is performed with both static and dynamic methods.

**5-2-1- Static loading in shear walls**

The static method works with two loading mechanisms. For the first loading mechanism, the force is applied uniformly (or monotonically), and in the second scheme, it is applied in a cyclic way (Figure 8). In both methods, loading can be either displacement controlled or force controlled. In some cases, both phases of force-controlled and displacement-controlled are utilized simultaneously. For example, in the first phase, the wall is loaded up to the point of yielding of the longitudinal bars (or the first crack at the cross-section of the walls)—as controlled by force, and then the structure is continued to be pushed using a displacement-controlled loading protocol.

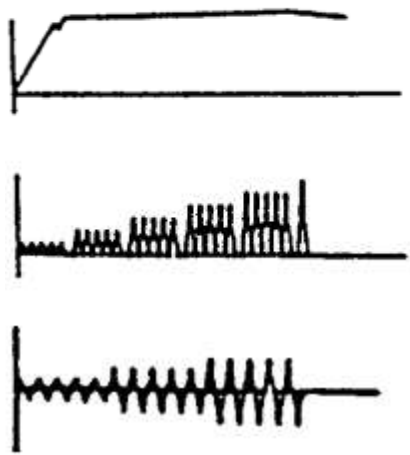

Figure 8. From top to bottom, uniform loading pattern; one-way cyclic loading pattern; two-way loading pattern [10]

In general, different loading regimes are used for both of the aforementioned methods. As stated earlier, cyclic loading can be used for both displacement-controlled and force-controlled loading. In both of the mentioned methods, there is a high variation in the shape of the loading pattern—each with different intensities—though all of them claim to represent earthquakes of a particular intensity. In the force-controlled loading protocol, the rate of load increments can be both constant and variable. In displacement-controlled loading mode, both the number of loading periods and the loading increments (displacement increments) vary. So there are different patterns for cyclic loading for both methods. This difference affects the results of the experiments and this issue (using several models for cyclic loading) has been investigated as an independent variable in the study of several researchers [13].

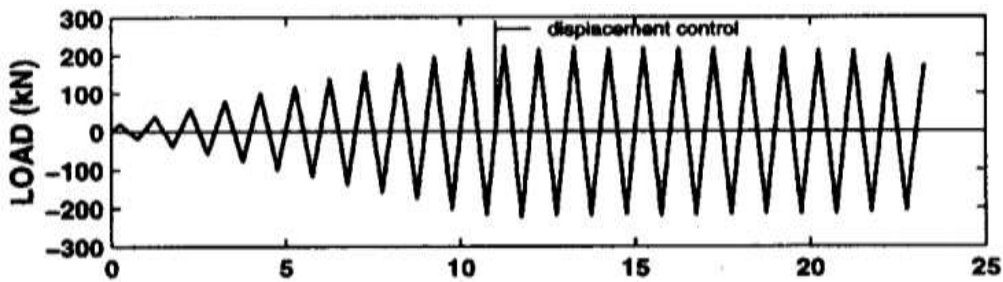

Figure 9. Example of a lateral loading regime that is designed through a combination of force-controlled and displacement-controlled phases [13]

5-2-2- Dynamic loading on the RC shear walls

Shear walls are usually subjected to dynamic loading using shaking tables. Vertical loading is applied using the lab equipment that comes in the following pages. Seismic masses are attached and mounted to the shear wall using a side retaining structure that will only work if the vibration is sensed at the foot of the structure. In this case, the vertical weight resulted from these masses is carried by the lateral retraining structure, and the bearing through which the mentioned weight is transferred has been attempted to have the least friction with the RC walls (Figure 10) [3]. Therefore, seismic masses are applied to the RC shear wall in such a way that it does not affect the vertical load of the wall. The acceleration record can be used as dynamic inputs at the level of wall support by a variety of methods. These methods can include scaling the acceleration record according to a specific acceleration design spectrum [3] or to scale an acceleration record incrementally [6]. Selecting an appropriate acceleration function for the research work itself is always an engineering challenge because the results of the experiments are so sensitive to the adopted record selection procedure [11].

In static loading and in both cyclic and uniform methods, the functions of different loading patterns represent conditions that may occur in a predetermined (e.g., moderate or strong) earthquake. The first criticism about the loading patterns is that the differences in the shape of these functions will themselves cause a large dispersion in the experimental results collected. Thus, finding which loading regime would give the best estimate is seriously under question.

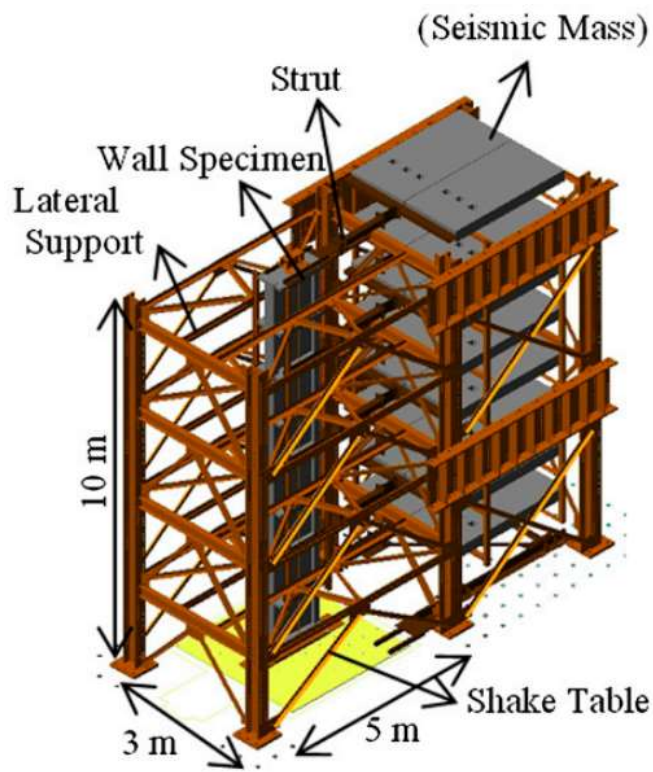

Figure 10. An RC shear wall which is subjected to a dynamic lateral loading [11]

In one research work [5], scholars engaged in this study constructed and considered two RC shear walls with the same geometrical and mechanical properties and subjected them to two different types of static and dynamic loading. The purpose of this work was to investigate the sensitivity of the test results in loading regimes. The results are as follows:

1. The maximum (peak) strength of the two examined RC walls is different from each other

2. The displacement cyclic curve obtained from dynamic loading has a stiffer manner than the same curve calculated from static loading

3. The failure mechanisms in the two walls are different

The article states that cyclic static load is a good representative for dynamic loading, but if we consider the above explanations, this does not seem to be the case. It can be apparently understandable that the best method is to use a large scale shaking table test to find the real nonlinear behavior of RC shear walls [3].

## 6- Independent failure mechanisms in RC shear walls

A prerequisite for the design of ductile structural walls is that flexural yielding at the vicinity of the plastic hinge defined at the cross-section of the wall should control the strength level, non-elastic deformation and thus energy dissipation throughout the structural system. Therefore, as a matter of principle, any brittle failure mechanisms or even those with limited plasticity should be avoided. To achieve this goal, one can use the capacity-based design method for finding the general configuration of the structural member and continue to complete this mission by using appropriate reinforcement details for the plastic hinge area.

The main source of energy dissipation in shear walls should be the yielding of bending or longitudinal bars within the area of plastic hinges (usually below the shear walls) [1, 13]. These mechanisms are illustrated in Figures 11 (a) to (e). Damage modes that must be prevented from occurring in RC structural elements—including RC shear walls—are: diagonal tension (Figure 11) (web failure), diagonal compression due to the excessive shear demand in the shear walls (web failure) (Figure 12), buckling of the thin sections of the shear walls or the same buckling mode that may be witnessed in the reinforcement bars located within the compression zone of the walls (Figure 12), shear slip failure along the construction joints (Figure 11), shear failure and bond-slip failure along the overlapping reinforcements [1].

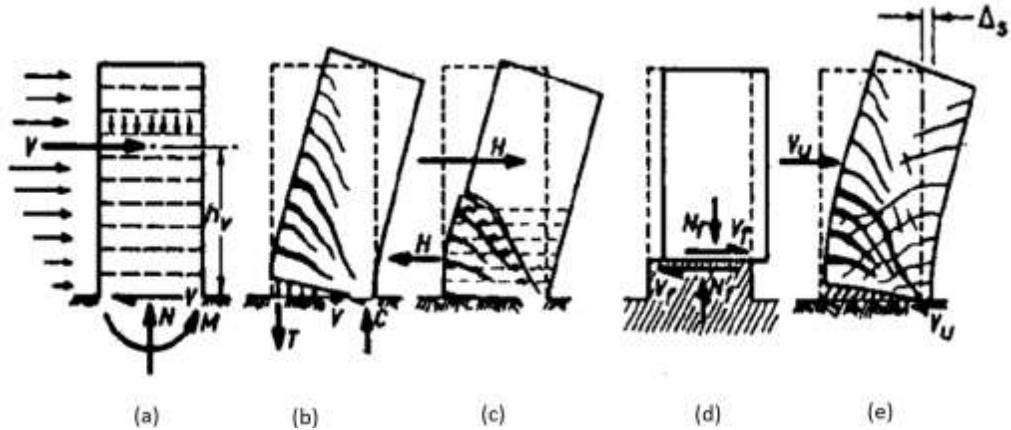

Figure 11. Independent failure mechanisms of RC shear walls [1]

As can be seen in the figure above, in order to see a particular failure mechanism in general, a specific cracking pattern in the RC wall must be formed in advance. This pattern has a major impact on how the wall gets damaged or how its energy dissipation mechanism would emerge. In the following sections, we will talk more about the cracks in RC shear walls.

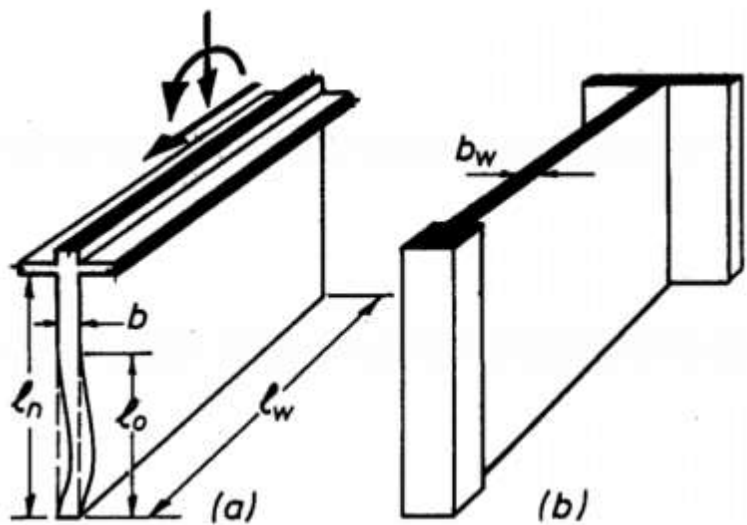

Figure 12. Out-of-plane buckling of the RC shear wall [13]

In the following figure (Figure 13), the force-displacement curve of a shear wall that is being controlled by shear is illustrated. As can be seen, the wall has a severe progressive decrease for both stiffness and strength levels. There is also a pinching in the center of the graph. In general, this wall has very little energy absorption. In contrast, Figure 14 shows a cyclic curve of a wall that behaves much better because it has been designed using capacity-based design principles [1].

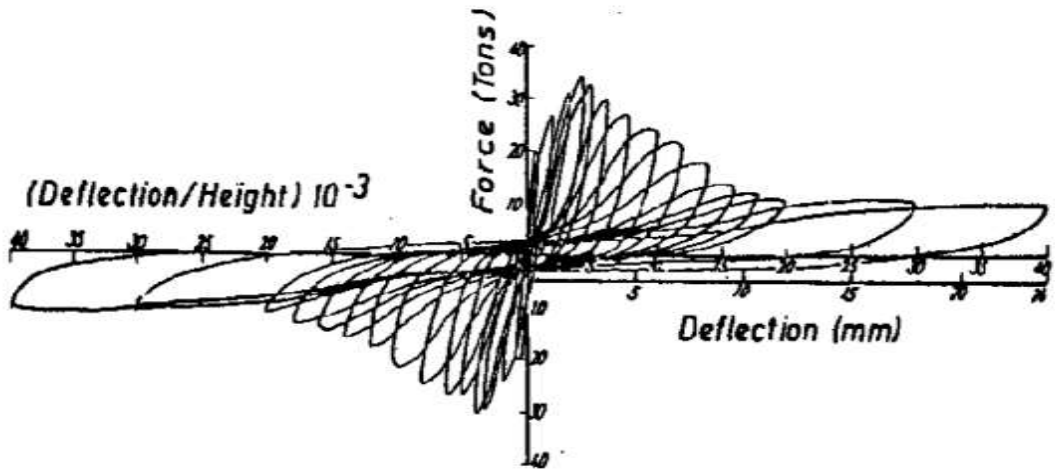

Figure 13. Force-displacement curve of RC shear walls with a shear behavior [1]

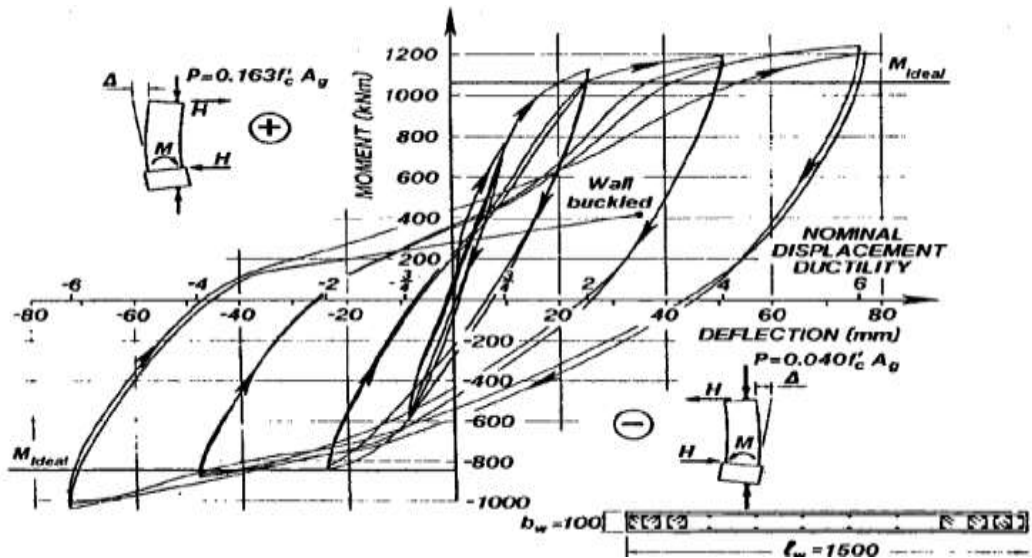

Figure 13. Force-displacement curve of RC shear walls with a bending manner [1]

# 7- Cracks in shear walls

The forces that are actually applied to structures from severe earthquakes are much larger than the values that the current seismic codes report as equivalent static loads for the linear analysis and design purposes, and the design of structural systems rely on elastic behavior of such systems in earthquakes; except for some specific structures (hospitals or special structures), there is no economic justification and the resulting sections would be greatly enlarged in case we use elastic methods for the design purposes. Thus, in structures, including shear walls, nonlinear deformation will definitely occur in a moderate or severe earthquake, so various cracks may also occur afterward. It is mainly difficult to observe cracks due to vertical loading in the walls [3], and this matter may be the main reason of the difference between the theoretical and real cracking load on the shear wall element because in the laboratory environment it is very difficult to observe cracks that are very fine [4]. In the following sections of this report, we will talk about the different types of cracks and their properties:

## 7-1- Horizontal (or bending) cracks

Horizontal cracks are of bending type. These cracks usually occur in the boundary element of the wall. The length and density of these cracks at the foot of the shear wall is greater than the ones on can find at other parts of the wall. As we move upward from the base of the shear wall, both the crack length and its density and thickness (i.e., the width of opened cracks) will decrease [4]. In walls with bending behavior, flexural deformation contributes to a major part of the wall deformation (the crack number 1 in Figure 14 (a)). (For learning more about the shear wall deformations, please refer to the section provided for deformation of shear walls

## 7-2- Inclined (or shear) cracks

Inclined cracks are of shear type. These cracks usually occur at the shear walls with an angle of about 45 degrees [7]. Contrary to what we had for horizontal cracks, the number and density of inclined cracks at the height of the wall cannot be recognized and detected with reasonable certainty (the crack number 2 in Figure 14 (a)).

### 7-3- Horizontal-inclined (or bending-shear) cracks

The horizontal-inclined cracks are actually a combination of crack types we discussed before—horizontal and inclined cracks. Horizontal cracks first form at the base of the wall, and then the shear cracks start appearing at the bottom. Usually, the number and density of these cracks at the base of the wall are greater, and the slope of these cracks increases with the distance from the base of the wall. So in this case, the cracks will rotate and we may have crack rotation from bottom to the top point of the wall [4]. These cracks at the base of the wall are sometimes very deep and so long, and they continue to appear across the RC wall. Also, these cracks at the base of the wall can itself cause shear distortion at the bottom of the wall (the crack number 3 in Figure 14 (a)).

### 7-4- Vertical cracks

Vertical cracks are usually of the compression type. In the claws of the wall, vertical cracks usually occur prior to the concrete cracking in this area due to high pressure originating from compression forces of this zone. As the strain of the concrete in this area approaches a value recognized as a concrete crashing strain, these cracks would be visible to the naked eyes [13]. In many cases, strains are either unrecorded or unacceptable if they tracked and recorded beforehand [3, 6]. This may be mainly because of the removal of the strain gauges between the formed cracks, which is due to the local bar yielding that occurs among the opened cracks [5].

### 7-5- Construction joint cracks

By applying a load to the shear wall, the construction joint can be converted to cracks under the wall. Unless this crack itself may not become an independent failure mechanism, it will cause shear sliding deformation at the base of the wall. It is worth mentioning that the sliding deformation will thus contribute to the top displacement of the RC wall.

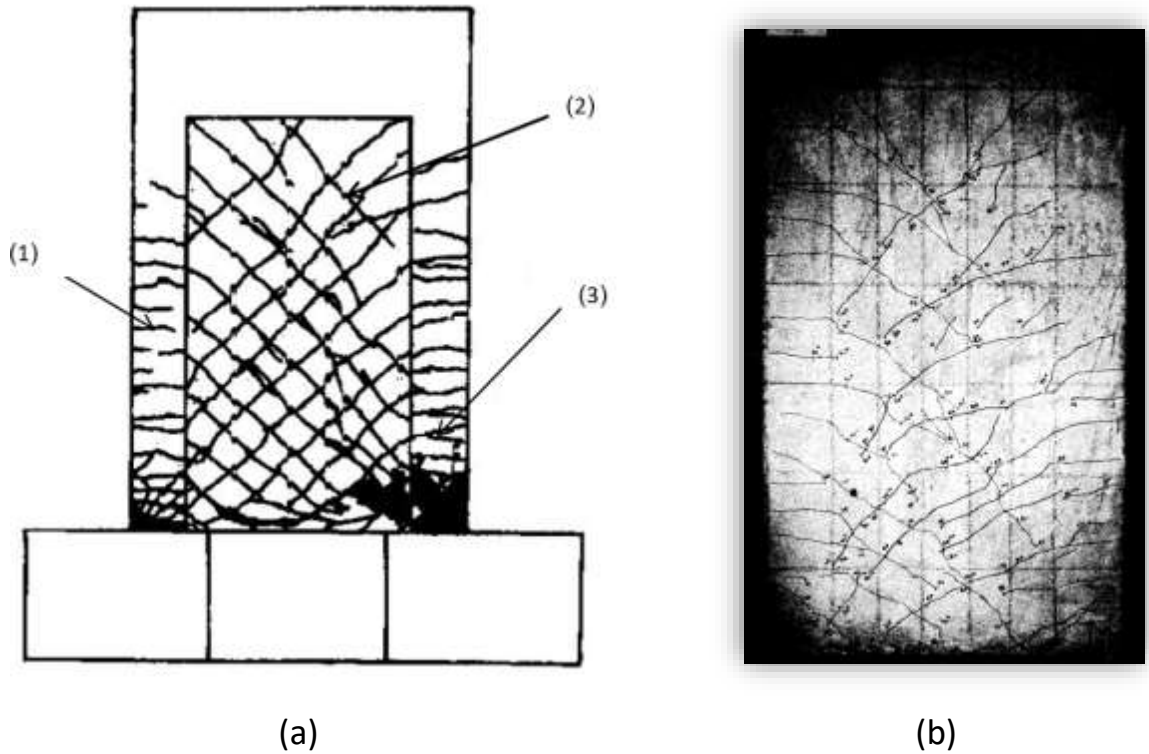

(a)            (b)

Figure 14. Patterns of cracking in the wall, and types of cracks in the RC wall [6]

Cracks are formed and appeared in a shear wall depending on different conditions such as the type of reinforcement (the type of reinforcement as well as the distance and the percentage of reinforcement formulated for a cross-section), the sample dimensions and various other parameters. Therefore, it should be noted that the cracking arrangement in a wall can itself lead to a failure mechanism. This failure mechanism, in general, and in accordance with the wall conditions, can be either independent (i.e. one of the mechanisms described in section 6) or a combination of independent failure mechanisms in the RC walls.

## 8- Sources of deformational demand in RC shear walls

Cracking in shear walls and yielding of longitudinal reinforcement may considerably reduce the stiffness and strength of the RC walls. As a result, sorts of deformation occur in the shear wall. The main sources of deformation in shear walls are as follows:

$$\Delta = \Delta_{flexure} + \Delta_{shear} + \Delta_{slid} + \Delta_{B.R} \qquad (6)$$

where,

$\Delta_{flexure}$: the deformation caused by bending in the shear walls
$\Delta_{shear}$: the shear deformation in RC shear walls
$\Delta_{slid}$: the shear-slip deformation at the base of shear walls
$\Delta_{B.R}$: the deformation caused by rotation of the support at the base of the shear walls

## 8-1- Bending deformation

The bending deformation contributes to a major part of the whole deformation of the shear walls (Figure 15). In shear walls which have a flexural-dominated seismic performance (walls with a dimension ratio of 2 or more), mainly more than 80% of the displacement on the walls is due to bending deformation [6] (Figure 16). In walls believed to have shear manner (walls with dimension ratio below 1), the amount of bending deformation is varied from 40 to 60% (Figure 17). In this group of the shear walls, the percentage of bending deformation is in the same range—i.e., from 40-60%—even when the RC wall lacks shear bars [11].

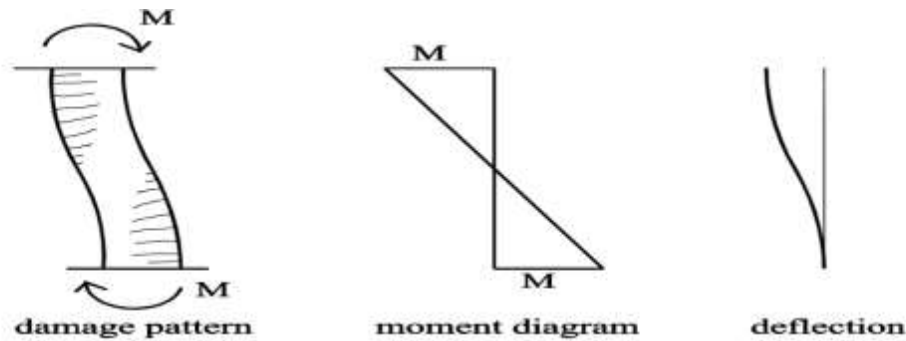

Figure 15. Bending deformation of an RC wall [11]

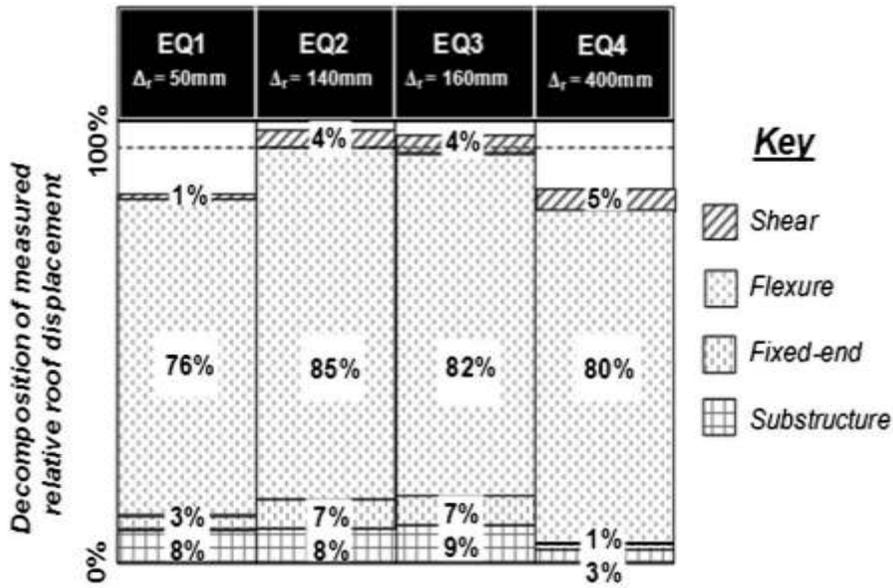

Figure 16. Percentage of bending, shear, rotational, shear-slip deformations in flexural-controlled RC walls [3]

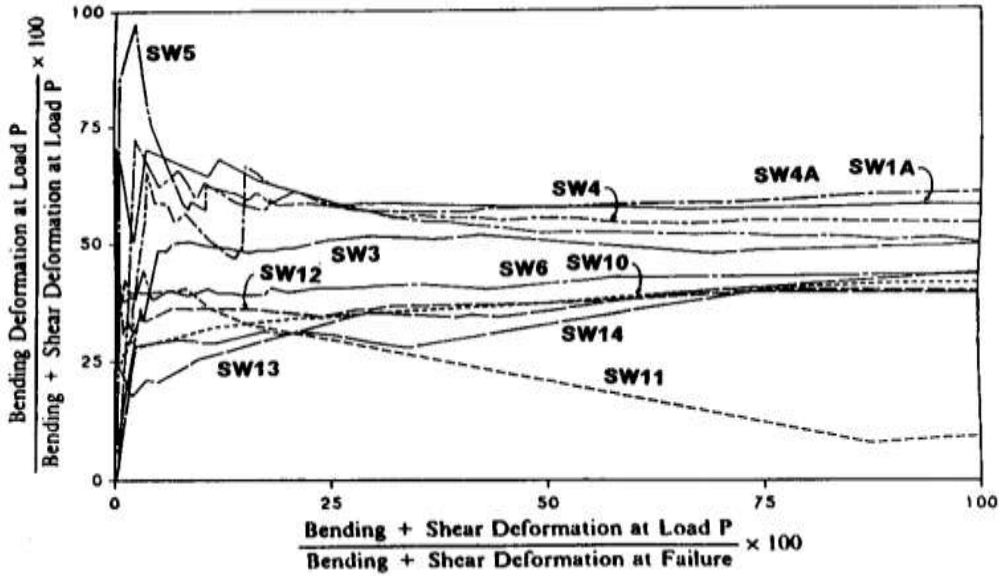

Figure 17. Percentage of flexural and shear deformations in a tested shear-controlled RC wall [11]

## 8-2- Shear deformation

The amount of shear deformation in the RC walls completely depends on the behavior of the wall subjected to lateral loads. This means that we can have a different percentage of shear deformation for two different RC shear walls, the squat and slender shear walls (Figure 18). In shear walls which have a flexural performance (walls with a dimension ratio of 2 or much more), usually less than 5% of displacement on the RC walls is due to shear deformation [6] (Figure 16). In walls with shear-dominated behavior (walls with dimension ratio below 1), the amount of shear-induced deformation varies from 40 to 60% (Figure 17). Therefore, the amount of shear deformation in the shear walls controlled by shear may be quite considerable.

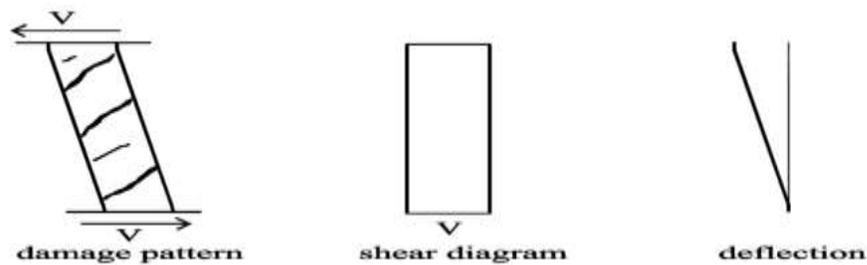

Figure 18. Shear deformation of an RC wall [11]

## 8-3- Shear-slip deformation

Shear-slip deformation usually appears in shear distortion manner—shear distortion or angle change. This type of deformation may occur at several points, and one of them is where construction joints exist. When the construction joint gets converted to a joint crack (see section 7 on the cracks of shear walls), shear-slip deformations will occur due to the weakening of this area of the wall [10]. The same kind of deformation happens at the location of the main flexural cracks too [12]. Occasionally, these deformations occur at the top of the wall. The reasons for this matter may be the yielding of longitudinal bars near the rigid slab above the wall, crushing of concrete on both sides of the wall, as well as the low-quality concrete at the top of the wall [5] (Figure 19).

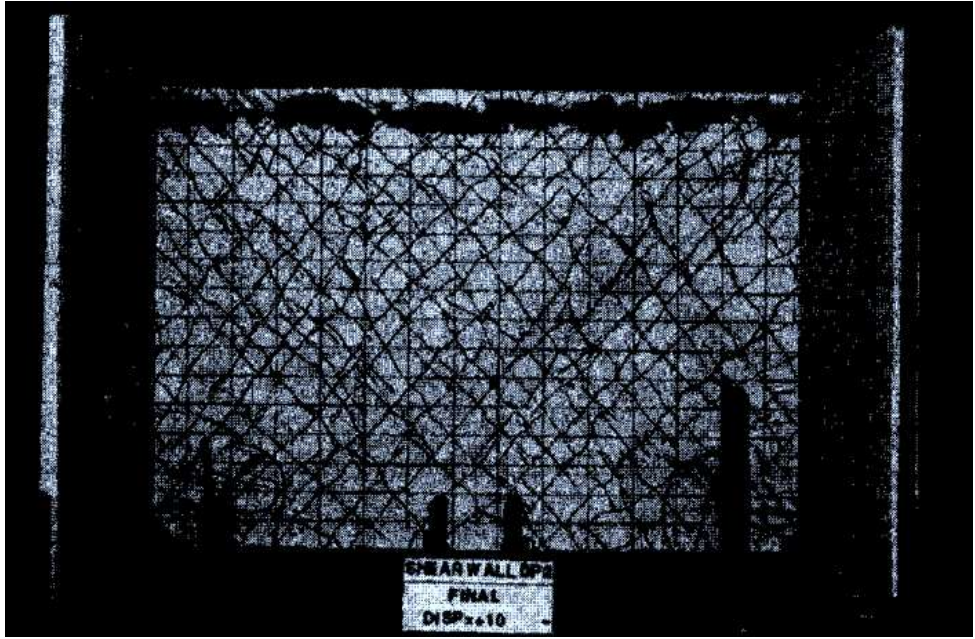

Figure 19. Shear-slip deformations at the top of the wall [5]

## 8-4- Rotation-induced deformation of the support

experiments have shown that relatively wide cracks are formed at the common surface of the wall and the foundation. Besides, this type of cracks can be also seen at the point where the rigid beam gets joined to the wall. As explained earlier, this rigid beam is considered as the rigid foundation of the RC wall. The formation of these wide bending cracks is mainly due to the penetration of axial strains to the full length of the tensile bars that are within the foundation or the joining point of the beam and wall. By increasing these longitudinal strains in the tensile rebar—which turns out to the formation of plastic strains along the longitudinal bars—and by the loss of adhesion between the steel and concrete around these bars, extending and sliding of reinforcement joints at the common surfaces of these joints can be of notable magnitude. The overlapping and slip of the tensile reinforcement in the foundation results in additional rotations at the bottom of the fixed supported point of the wall, which cannot be captured through a simple analysis that is normally based on the bending manner of the elements (Figure 20). During testing, displacement meters are used to read this type of rotation, where one of the reader legs of this displacement meter are placed one side of the wall while the other leg is put to the base of the wall. The amount of this rotation varies in different experiments. The percentage contribution of these deformations to the whole wall displacement has been found to be negligible (the maximum value reported in this case is 8% from the total wall displacement [6] (Figure 20)). In a number of research works reviewed, even the amount of this displacement has been so small that the researchers have declared that the assumption that the foot of the wall is completely fixed is quite correct [2, 13].

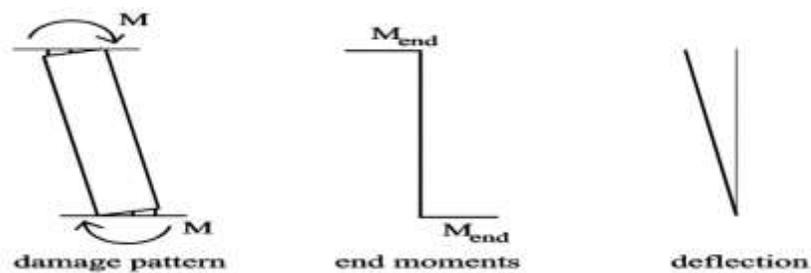

Figure 20. Rotational deformation at the base of the wall, which is due to the rotation occurring in the support [13]

## 9- Observed damages and definition of failures

According to the force-displacement curve of an RC member, one can readily find general information about the seismic characteristics of the member being considered. But it cannot be found that a point (a regular pair on the curve) of the load-deformation curve corresponds to what event in the shear wall element theoretically. For this reason, in tests performed, the test is stopped and the level of damage (Damage deformation, $\Delta_{damage}$) is recorded once a major event occurs (Figure 21).

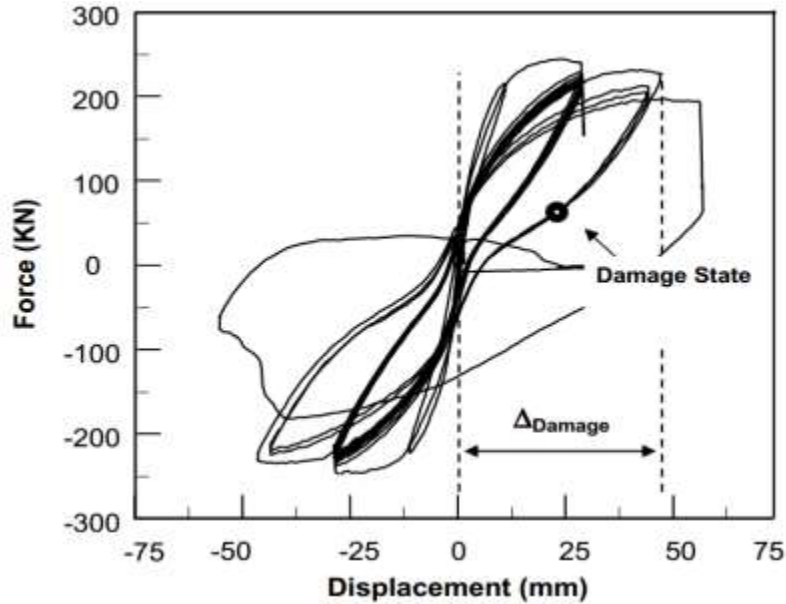

Figure 21. Displacement related to a specific level of deformation that is associated with a damage state [4]

The levels associated with the damage deformation are usually reported for the following situations:

• The onset of concrete spalling in the paw of the walls, it can be defined as the first observation of concrete scaling.

• Initiation of the severe concrete spalling, concrete spalling in this case would cover a considerable height of the concrete within the regions associated with the boundary elements of the wall. This spalling increases along the wall on both sides toward the center. The amount of this concrete spalling is related to the failure definition of the experiment being conducted in a laboratory. We will present the definitions of failure (fracture) in the following parts of this technical report.

• The buckling of longitudinal reinforcements in the RC wall, it is defined as the first sign of buckling in the longitudinal rebar.

• Fracture of the longitudinal bars in the RC wall, this can be considered as the first sign of fracture of longitudinal reinforcements.

• Fracture in the transverse reinforcements of the wall, this is also defined as the first observation of the fracture in transverse reinforcements or the opening of the joints of the steel stirrups.

• Out-of-plane buckling of shear walls, it should be defined as observing of the first sign of out-of-plane buckling of the wall.

As stated up to here in this section, you may see different definitions for the expected levels of damage. For each test, usually, a number of them can be seen. In general, the test usually proceeds to the point where the net strength of the element drops to 75 to 80 percent of peak strength. Then, and after reaching this performance point, the amount of displacement associated with this dropping point would be meticulously recorded. In fact, the concept and definition of failure are of relative quality indeed. As can be seen so far, there are different levels of failure, where each of them can be a form of failure. In experimental tests, the test procedure would not proceed until the wall is completely devastated and gets collapsed (in our search range, no experiment with this type of complete collapse has been found). The test will either traditionally proceed up to a point where a 75-80% drop in peak strength is detected [2, 4, 10] or it would be stopped if the concrete spalling mode of failure is observed in the test specimen—i.e., the spalling failure mode at the edge of the claw or part of the claw in the RC shear wall [4, 10, 13, 21]. In several experiments, this spalling failure mode has been found to be correlated with a strength level that is less than 50% of recorded peak strength [2, 5].

## 10- Learning insights from failure to the point of fracture in the RC walls

In examining the expected levels of damage in the RC walls up to a point once a type of failure defined in various experiments occurs, interesting events that are worthwhile to follow would happen. By studying this process, we can investigate the effect of different factors on the nonlinear behavior of shear walls. As already mentioned, the shear walls are divided into two groups according to their bending and shear performance. Perhaps the most important criterion for identifying a wall's performance is the type of design method as well as the geometry and properties of the considered RC wall. In this section, the aspect ratio criterion is considered for separating the shear and bending performance of the walls.

## 10-1- Shear walls with flexural performance

Damage assessment of bending-dominated shear walls can be investigated in two stages; in the first stage we will evaluate the performance of the shear walls before yielding the main bending bars, and in the second stage we will evaluate its seismic performance after the yielding phase of longitudinal rebar. The behavior of these walls is usually similar to each other prior to the yielding of the main bending bars and is the same for all RC walls recognized to be in this category. In the post-yielding phase, the behavior of RC walls in this category would usually vary depending on its exclusive properties.

### 10-1-1- Pre-yielding behavior

As mentioned earlier, the structural performance and pattern of wall cracking before the yielding phase of the main bars are usually the same for all RC walls of this group [4, 13] (Figure 22). At the first stage, the wall usually experiences horizontal cracks first, and then these cracks are usually distributed at the height of the wall. In the lower part of the wall, these cracks are denser and longer than the upper part of the wall [4]. When more loads are applied to the wall, these cracks (shear-bending cracks) penetrate the wall at an inclined angle. The slope of the inclined shear cracks is steeper in the upper part of the wall. These cracks usually appear with small widths at this stage and act in an elastic manner when the walls get unloaded [4].

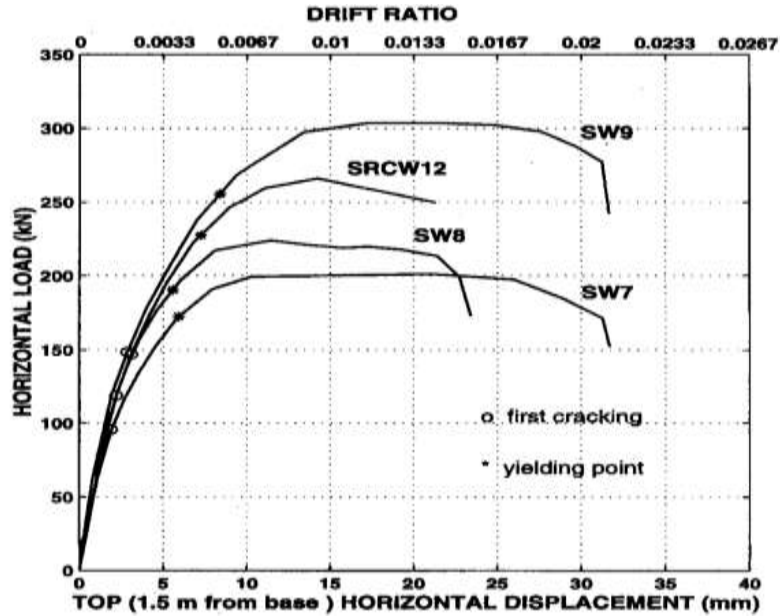

Figure 22. Pre-yielding behavior in a flexural-dominated RC shear walls [4]

## 10-1-2- Post-yielding behavior

After yielding of the main bending bars, the growth rate associated with the number and width of bending-shear cracks increases more than the other cracks that normally appear in this kind of shear walls. Next, we will have the initiation of the web inclined cracks. As the external load on the wall increases, the number of these cracks increases also. This cracking pattern then continues with greater speed and growth at the foot of the wall. From then on, the growth and the number of cracks will remain constant, and only a very small number of cracks would occur among the major cracks already existed [4]. In this group of shear walls, the energy dissipation mechanism is controlled by the plastic strains of the longitudinal bending bars. It is of worth to mention than these plastic strains are released after the yielding of the flexural rebars. Therefore, the growth and development of new cracks are not the main reason for the energy dissipation of such walls [13]. These aforementioned inclined cracks then cut each other and form grids with a criss-crossing pattern on the wall. Due to the high density of these cracks at the foot of the wall, the size of the developed grids is very small. While the concrete at the claw of the walls is compressed under high pressure, the concrete cover of the shear wall element would be heavily destroyed in this situation. The spalling of the concrete in the claw zone usually occurs once vertical cracking has emerged in this area [4, 13]. After this stage, usually, the process of cracking and spalling of the concrete will continue on both sides of the sections that are related to the concrete of the claw area. In these extremes points of the RC walls—claw zones—the main steel bars will buckle altogether subsequently. These events, stated up to here, decrease the strength level of the walls. But the experiments are typically terminated before this step (Figure 23).

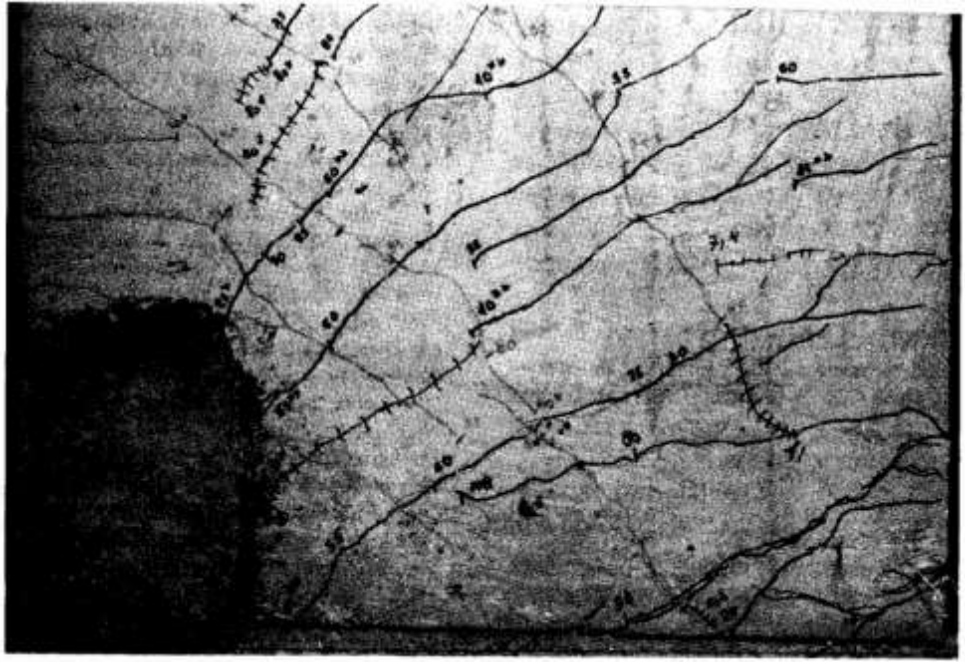

Figure 23. Crushed concrete in the claw of an RC shear wall [13]

In the post-yielding phase of the walls, the wall's performance is usually influenced by the following factors:

**Vertical load**: Vertical load on the wall reduces the main tensile stress on the RC wall, which reduces the number of diameter cracks in the web of the shear wall [4]. The vertical load on the wall also controls the width of the cracks, which means less distortion and also less pinching in the seismic performance such walls [7]. RC walls with a vertical load in this group are stiffer (the secant stiffness) and more strength (3 to 6% more strength). This is due to the further confinement of these types of walls, which is caused by the presence of the vertical loads in this case.

**Shear-compression ratio**: The mathematical definition of the shear-compression load ratio is given in Section 2 which is on the general characteristics of the shear walls. The higher the ratio of shear walls, the denser the criss-crossing grid pattern density. In addition, the higher the shear-compression load ratio is in the shear walls, the lower the stiffness reduction of the walls would be in the loading phase [4].

**Earthquake intensity**: In this group of the walls, shear cracks are added to the bending cracks once the earthquake intensity increases [6]. However, a more in-depth studies in this case shows that the cracking pattern in the shear walls does not change with increasing earthquake intensity [3, 6] (Figure 24).

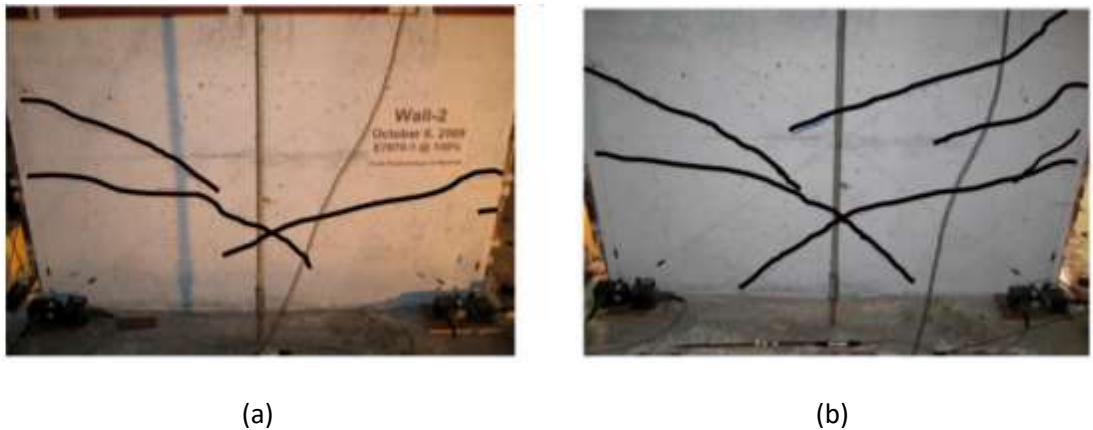

(a)          (b)

Figure 24. With an increase in the magnitude of an earthquake from left (a) to the right (b) in this figure, the pattern of wall cracking remains constant, with only a few shear cracks being appeared in figure (b) [6]

### 10-2- Shear walls with shear behavior

In shear walls with shear manner, the corresponding point of yielding in the main flexural bars—witnessed in RC shear walls with bending behavior—is the occurrence of the first diagonal crack in the wall. Prior to this point, the amount of stiffness reduction in the walls is negligible, and no considerable event happens in this case. After this point, the process in charge of nonlinear behavior of the walls would be started and many other events associated with different damage states would happen correspondingly. Various factors will actually affect these events. So, in the following parts of this section, we investigate the behavior of these walls in two stages—before the first diagonal crack and after the first diagonal crack. In the behavior of the walls after the first diagonal crack, we examine a series of influential events on these walls.

**10-2-1- Before the first diagonal crack**

As mentioned before, performance and pattern of wall cracking before the first diagonal crack are usually the same for all walls in this group [11]. At this stage, the wall first experiences very small horizontal cracks. These cracks have very little effect on wall stiffness and can usually be ignored to reduce the wall stiffness. When the loading proceeds to be applied on these walls, the stiffness of the walls decreases dramatically after the first diagonal crack in the wall [5].

**10-2-2- After the first diagonal crack**

When the first diagonal crack in the shear wall occurs, the lateral stiffness of the shear walls sharply increases, and then displacement demand in the wall will be subsequently intensified. Next, the shear-bending cracks appear in the wall and thereafter the yielding of the main wall rebars occurs (in some references shear-bending cracks have been reported to appear before the wall diagonal crack [10]). Unlike shear walls with bending behavior—where wall shear displacement is only a function of bending deformation—in shear-dominated RC walls, shear deformation is a significant part of the resulted yield displacement [12]. After this step, the wall will reach its bending capacity at its peak strength (the shear that is equivalent to the bending strength of the element). Once the wall member reaches this strength level, the main bending cracks occur at the bottom of the wall. After this damage state, since the strength level of the member would decrease to the shear-slip resistance of the wall's cross-section (Figure 25), significant shear distortion deformations would emerge along these cracks as well (Figure 26). In general, these walls reach the point of failure in several ways. At the failure point, modes of failure are usually divided into two types of damages that are concentrated on the web of the wall or around the claw of the shear-dominated RC wall.

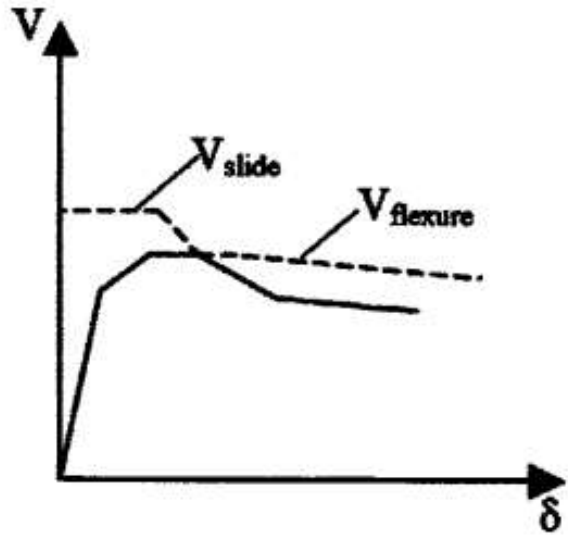

Figure 25. Behavior curves of squat shear walls and loss of its strength capacity to the corresponding strength standing for shear-slip mechanism of the wall [12]

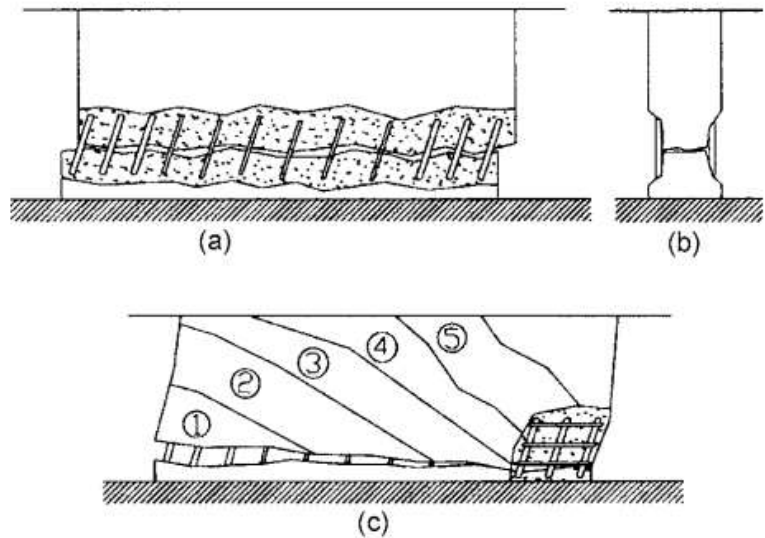

Figure 26. The main flexural crack and the distortion deformation at the foot of the RC wall [12]

The damages that occur at the web of the wall—which pushes the wall to a failure point—are also divided into several categories. Usually, the shear failure in this type of shear walls is either of tensile-based or compressive-based failure. If we consider the shear walls as a steel truss that has two diagonal axial members and imagine it is being loaded, one of these diagonal members would be subjected to a tension force while the other would be under a compression force [22]. When the external load in the system reaches a zero value, the cracks would not be closed yet. And once cracks are closed, only the compressive member is capable of conveying considerable amounts of the new loads during the reloading phase. In such a case, the stress of the concrete increases in the compressive diameter of the wall and exceeds the amount of stress that can be tolerated by the concrete. The intensified stress status of the compressive diameter may eventually cause damage to the wall or its claw. The tensile diameter member, in this case, will usually convey a small portion of the external shear force applied to the wall. Instead of the tensile diameter, the transverse bars take the responsibility of conveying the forces that should be gained by the imaginary tensile diameter. If the transverse bars get yielded, tensile failure at the web of the wall occurs.

There is also a specific case of damages seen in the web of the shear-dominated RC walls, which is reported by some references. In such cases, damage to the wall's web has been attributed to the uneven closing of the cracks, which is mainly due to the stiff boundary elements as well as the high value of axial loads. The area on which this type of damage is extended and propagated is in a rectangular shape (Figure 27). The effect of each of the mentioned factors would be separately discussed.

When a horizontal crack (the main bending crack) occurs at the bottom of the wall, many distortional deformations are developed in this area of the RC wall. At this stage, the wall strength or the shear capacity of it decreases to a strength level that is associated with the shear-slip failure mechanism. Next, the abrasion of the concrete edges come about within the main horizontal crack area. In this type of compression fracture, usually the toe area is severely damaged [12].

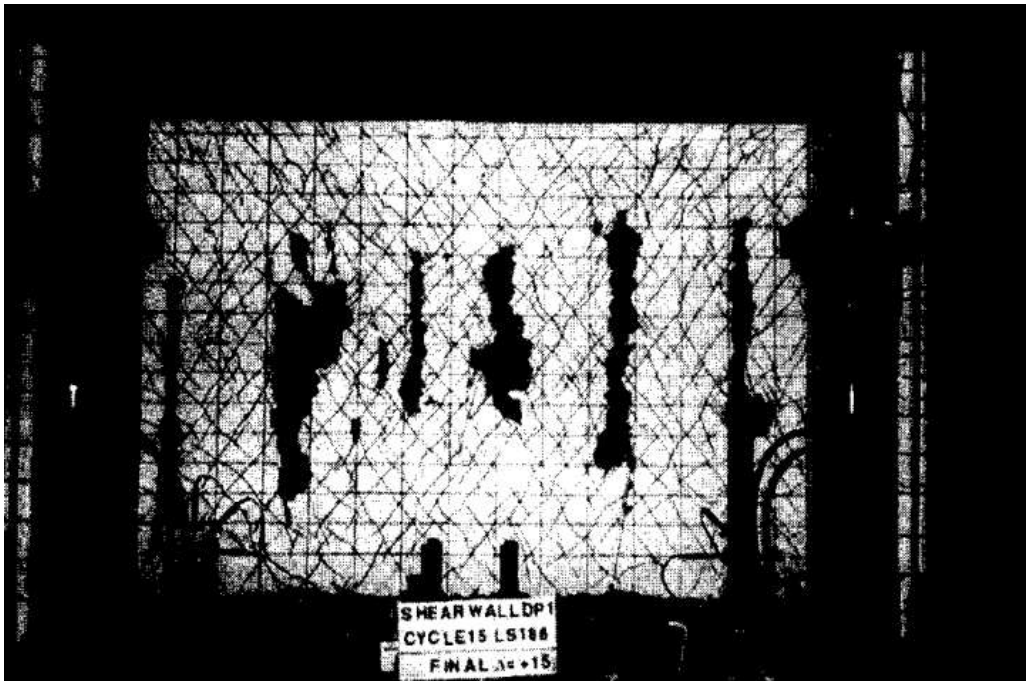

Figure 27. Vertical rectangular damages and failures in the web of RC shear wall [5]

For the behavior of the shear-dominated RC walls after the first diagonal crack, the wall's performance is usually influenced by the following factors:

**Vertical load**: In squat or short shear walls that have shear performance, pinching is apparently recognizable in their seismic behavior that can be readily evaluated by the load-displacement curves. The pinching in the seismic behavior of the walls is mainly a function of the rebars sliding and detaching from the concrete (loss of adhesion between rebar and concrete), or shear distortion that can be seen at the foot of the wall in such situations. [10]. Vertical load in addition to an increasing effect on the strength of cross-section of the wall element—which is also present in shear walls with bending performance—can also control and improve the issues mentioned above. Therefore, the vertical load on shear walls with shear behavior increases the strength capacity of the wall element, which can improve the seismic performance of the squat shear walls.

**Boundary Elements**: Boundary elements increase strength and shear capacity of the squat shear walls because they can make the core area of the wall more confined [10]. Also, these boundary zones can reportedly limit the width of the cracks found on the surface of the RC shear walls. The boundary element in the shear wall similarly makes the wall stiffer against the lateral loading. This causes the width of the diagonal cracks in the shear wall to be controlled, so by reducing the width of these cracks, shear distortion and shear deformation are reduced [7]. The stiffer the boundary elements, the more brittle the squat shear walls behave in general. It is worthwhile to say that it has also been reported in an investigation that focused on the shear wall with bending behavior [2]. In this investigation, the main bending bars were concentrated on both sides of the boundary elements in RC walls and this arrangement of the flexural bars results in brittle shear behavior and shear failure instead of the bending behavior and non-brittle mode of failure in this wall that one would expect to see from a wall with larger aspect ratio [2].

**Shear bars**: As you already know, shear bars are divided into two categories including the conventional-based and diameter-based formats. In the first case, shear bars are used horizontally and in the form of stirrups. In the latter case, these bars are used in an X-shape (or crisscrossing form) within shear walls.

**10-2-3- Pros and cons of diagonal bars in squat shear walls**

i. **Advantages**

Reduction in the number of rebars: According to the research published through the reference of [10], the use of diagonal shear bars reduces the total essential number of the rebars used in the shear wall. Therefore, the use of diagonal bars in shear walls would be more economical than conventional shear bars.

Crack width control: Given the fact that the shear-diagonal bars cut the diagonal cracks vertically, controlling and limiting the width of the cracks in this case would be much more desirable and efficient [10]. In this mechanism, the diagonal bars would work in direct tension form. In such a tension mechanism, the diagonal bars work with more energy dissipation. Also, limiting the width of the cracks due to the reduced amount of shear deformations in the wall itself results in better seismic performance of the shear walls (Figure 28).

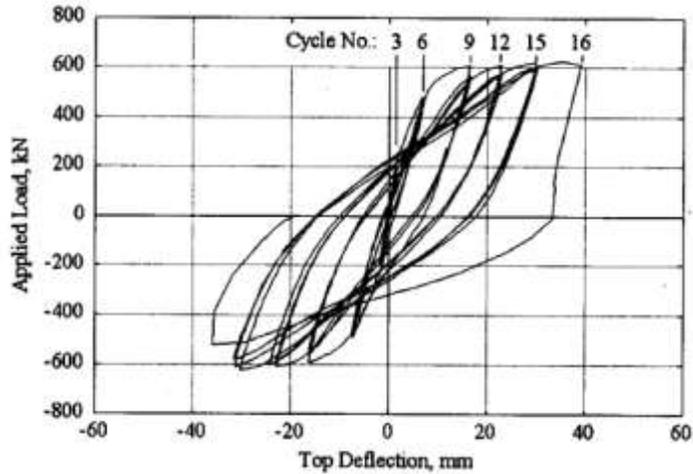

Figure 28. Improving seismic behavior of RC shear wall (with shear behavior) using diagonal bars [10]

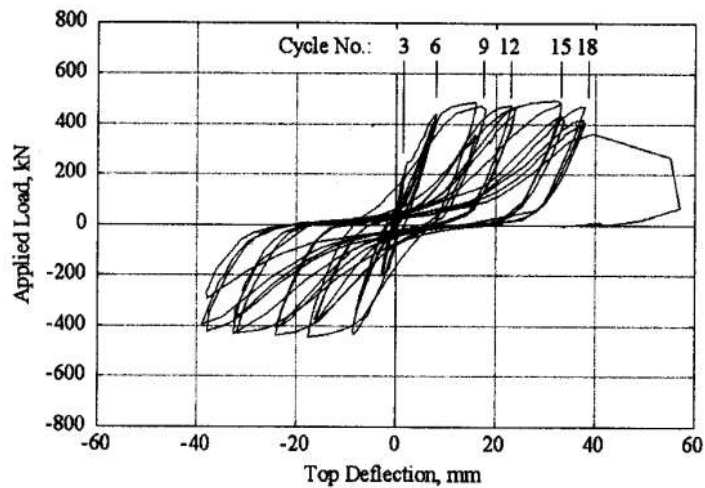

Figure 29. Seismic behavior of a wall depicted in Figure 28—with the shear behavior—but this time without the use of diagonal bars [10]

Reduction of shear distortion at the foot of the wall: In squat or short shear walls that have shear performance, pinching is apparently recognizable in their seismic behavior that can be readily evaluated by the load-displacement curves. The pinching in the seismic behavior of the walls is mainly a function of the bars sliding and detaching from the concrete (loss of adhesion between rebar and concrete), or shear distortion that can be seen at the foot of the wall in such situations [10]. The presence of shear-diagonal bars—due to its diminishing effects on the shear distortion at the foot of the wall—would reduce pinching that is normally seen in the seismic behavior of the squat RC walls (Figure 29). This matter can make the seismic performance of such walls become closer to the desired behavior of the shear wall members with bending behavior [10, 21].

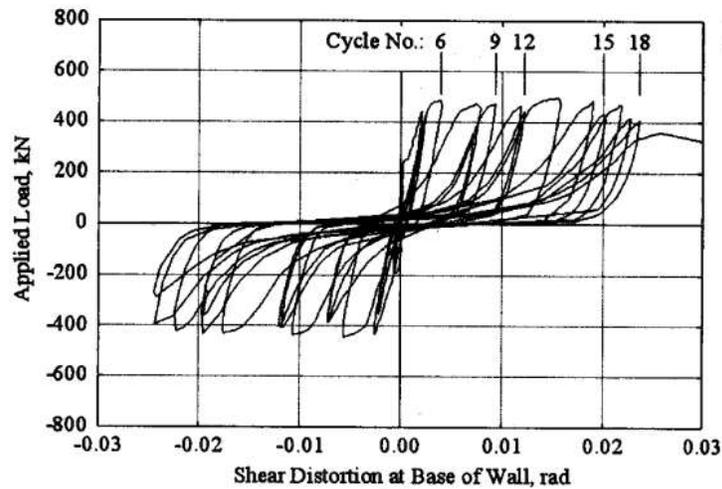

Figure 29. Shear distortion in a short shear wall without diagonal bars [21]

In the above force-displacement curve, shear bars are inserted into the RC wall to inhibit shear distortion that can be seen at the foot of the wall, but the final shape of the wall cyclic curve for shear distortion is as follows (Figure 30).

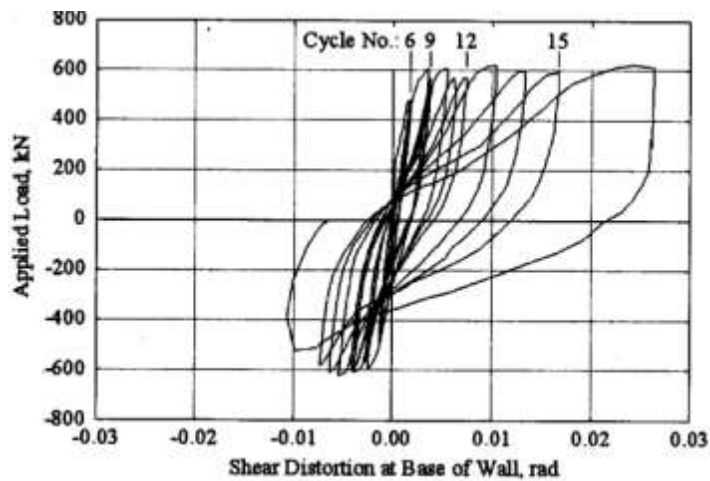

Figure 30. Cyclic curve of shear distortion versus applied force on walls armed with diagonal shear bars [21]

The reasons for the success of diagonal shear bars in inhibiting shear distortion at the foot of the wall can be related to the shear transfer mechanism along the height of the shear wall. As mentioned earlier, shear bars work in direct tension to prevent the occurrence of the diagonal cracks. In this load transfer mechanism, the amount of compressive stress on the other wall diameter—the hypothetical compressive diameter—is effectively reduced. In such a case, when the external load comes close to the zero value, unlike conventional squat shear wall systems with conventional reinforcing bars that cannot impart force in an effective way, in these short wall systems with shear-diagonal bars, bars working in tensile mode are capable of absorbing and transmitting forces. It is essential to note that such forces in conventional squat shear wall systems are not able to be reliably transmitted because open cracks cannot be closed when there are no such diagonal bars on the body of short shear wall elements.

<u>Improvemental effects on the shear strength</u>: The existence of bars placed in the wall in diameter form would increase the strength and shear capacity of the RC walls [21]. This increase in the strength level of the wall would be greater also if these bars are effectively kept away from the foot of the wall. However, the further the diagonal elements are away from the wall, the lower its function would be for the sliding control at the foot of the wall [10].

### ii. Disadvantages

<u>The difficulty for implementation</u>: It is difficult to use and implement diagonal bars in shear walls. This is not the case with conventional shear bars, and they are much easier to be implemented in RC shear walls.

Given the advantages offered for shear bars in shear walls, it seems that the difficulty in their implementation cannot be a factor to hinder us from using them in RC shear walls. In fact, the use of diagonal shear bars has many advantages that justify the difficulty in their implementation.

## 11- Field observations on shear wall structures in former earthquakes

Although the destructive earthquakes of the world have caused many irreversible financial and life damage, it should not be forgotten that an earthquake is a natural event in which the seismic performance of human-made structures are tested and verified by nature itself. In the meantime, hypotheses to construct such buildings are thoroughly challenged. Thus, given the nature of earthquakes and the reporting of technical teams following these events, one can be somewhat curious about the inherent performance of a particular structural system and be curious about a type of behavior shown in this case.

In evaluating the seismic performance of a structure, the overall performance of the structure is of particular importance. This is because, ultimately, the overall structural stability of a specific building is important during earthquake shocks. The overall performance of a structure has different levels and is a function of different factors that are related to the failure of various components of the structure. If some parts of a building are seriously damaged in an earthquake but the overall performance of that structure is satisfactory, then the overall performance and stability of the considered structure can be considered desirable and appropriate. In this section, we investigate the overall seismic performance of shear wall structures in previous real earthquakes.

The content of this section is collected using field reports prepared following a number of major earthquakes. Highly expert people are employed to write these reports (the reconnaissance reports). These reports provide information on the type and characteristics of the site—on which an earthquake record was received, the intensity and location of the event, as well as information on the performance of existing structures in an area stroke after an earthquake. This section uses the seismic performance of existing structures in the area, and among the reports for different structural systems, we only seek information related to the reinforced concrete structures with bending frames or the ones with the concrete shear walls.

### 11-1- The overall seismic performance of the shear wall structures

A number of buildings with shear walls can be seen in the reconnaissance reports, all of which share the same type of damages to structural components. In these buildings, often by forming short columns—generally short elements that include the wall itself—occurrence of the collapse in columns, which themselves play a very important role in the stability of structures, would be very probable. In these structural systems, this problem is often caused by the use of perforated shear walls, deep spandrel beams, and the parapet walls in the balconies of the evaluated structures (Figure 31). However, in the case of RC beams, this problem has been seen to occur when spandrel beam elements were used in the RC buildings reviewed (Figure 32).

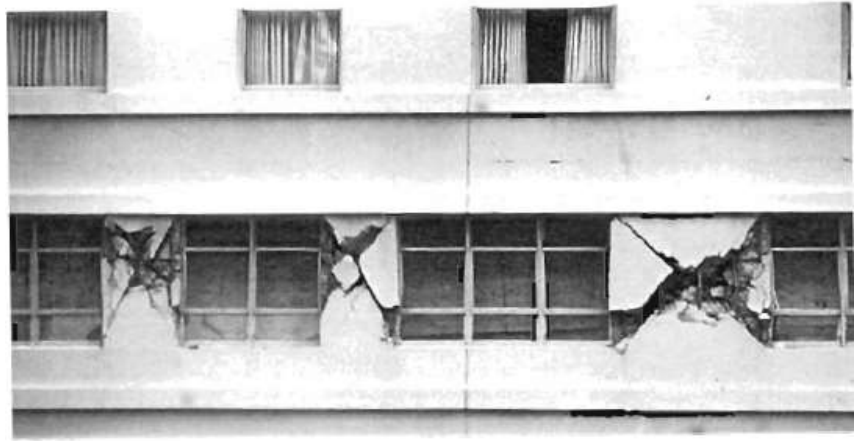

Figure 31. Damage imposed on the short columns and shear wall components in Northridge earthquake [23]

Due to the lack of element confinements and transverse bars required to provide proper ductility level—along with other factors that may cause this damage type, such as torsional forces in the floors—these structures may finally face damages because of these short RC members. The structural damages concentrated on these short elements can sometimes push the structural system to the level of irreversible partial collapse. Nonetheless, the decision to demolish the damaged buildings would be made by the owners of the structures. However, in all cases examined, it is obvious that these structures were able to drastically reduce inter-story drifts due to the high stiffness of the RC walls utilized in the system, thereby reducing structural and non-structural damages that could occur to other structural elements. Although shear walls themselves were damaged in these earthquakes, this did not cause the structure to lose its overall performance and functional ability. In all cases, reinforcing steel bars have maintained the overall integrity of RC shear walls and prevented them from getting crumpled [22].

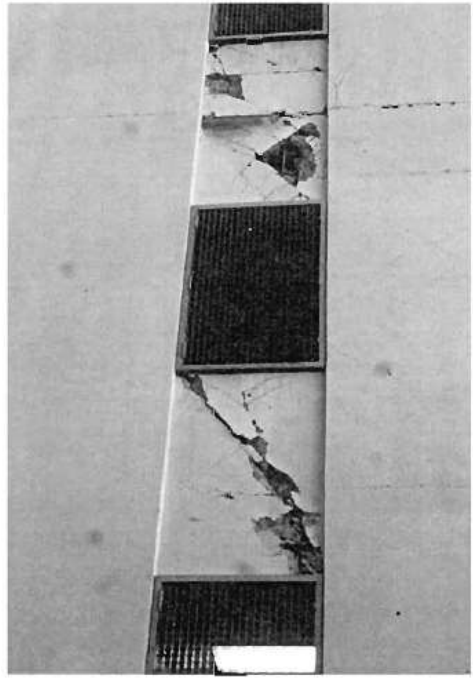

Figure 32. Failure of spandrel beams in Northridge earthquake [23]

In contrast to shear wall structures that were in the first category and have RC shear walls in such way that may cause the formation of short elements in the structure, there is a series of structural systems that get pronounced benefits from the RC shear walls in their lateral systems. In these structures, only superficial damages occurred after an earthquake of moderate to severe magnitude. For example, the structures represented the Indians Hills and Panarmo buildings—which both experienced the 1971 San Fernando earthquake—have suffered very little nonstructural damage. It is very interesting to report that these two structures named have also performed well in the 1994 Northridge earthquake. The Indians Hills and Panarmo buildings, which were equipped with tall RC shear walls, have both passed the aforementioned earthquakes with surface and no serious damage, respectively (Figure 33). It worth mentioning that these structures were built in areas where the worst structural damage were seen and reported. In sites near to them, a 10-story RC frame structure got completely collapsed while the other structures in near or adjacent places suffered severe non-structural damage.

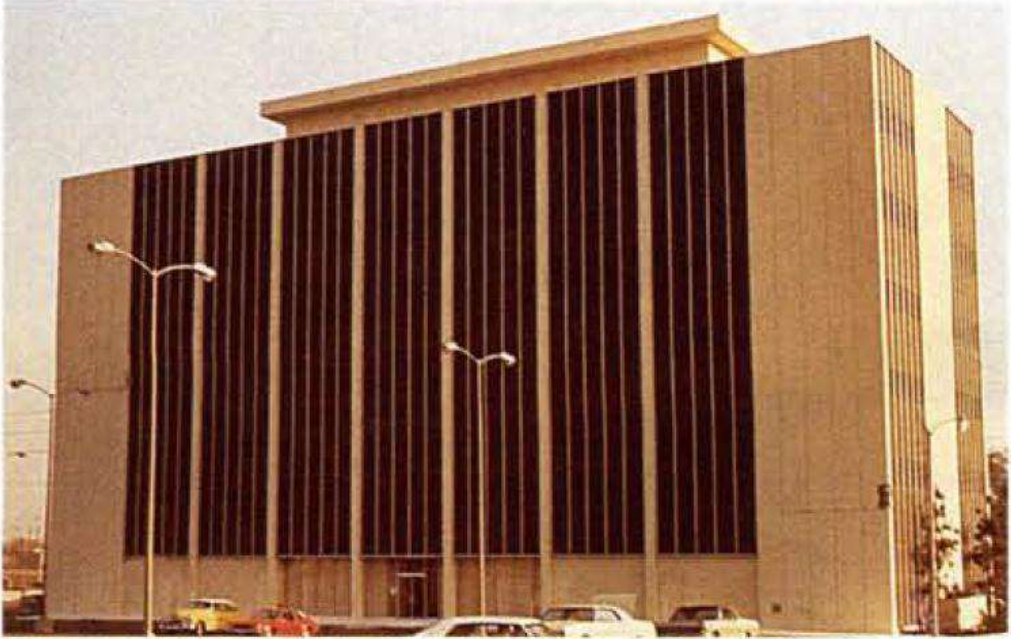

Figure 33. The Indiana Hills Building that reviewed in Northridge Earthquake [23]

# 12- Conclusion

As stated in the earlier section of this report, there are codes and views that hold this opinion that RC shear walls behave in a brittle manner. Consequently, a number of existing codes still suggest a lower ductility coefficient for designing of the RC shear wall structural systems than that they recommend for RC frame systems. To this end, we study the nonlinear behavior of shear walls in this technical report. In this case, experimental tests that were conducted on RC shear walls over the past 20 years have been used as the main source of this study.

In this study, we divided the shear walls into two types of shear walls—the one with bending performance and the other one with shear performance. In shear walls with bending behavior, usually desirable structural behavior with good seismic performance is expected. These members are usually considered to have appropriate behavior, and contrary to some opinions and seismic codes, they are not brittle members at all. Conversely, there are shear-controlled shear walls, which usually have poor seismic performance that leads to an inappropriate brittle failure. However, using tools such as diagonal bars and the other methods mentioned, one can make the performance of the squat walls get nearer to the optimum bending seismic performance that we should typically expect from a flexural-dominated RC shear wall. Therefore, in case shear wall members are going to be utilized as seismic lateral resisting systems, it should be noted that they should be designed to demonstrate a flexural manner. But if it is not possible, the wall elements should be designed in such a way that the shear wall system doesn't get damaged by shear and inappropriate brittle performance.

23. **A Blakeborough, P A Merriman, M S Williams.**, 1994, *THE NORTHRIDGE, CALIFORNIA EARTHQUAKE 1994.* London : Earthquake Engineering Field Investigation Team.

24. **Harati M**, *2014, Seismic performance of RC shear wall structures, M.Sc. thesis, Sharif Univ, Tehran, Iran. Doi:* 10.13140/RG.2.2.15402.82888

25. **Moghaddam H, Harati M**, 2015, *Seismic performance comparison of mid-rise moment resisting frame and shear wall system*, 7th International Conference on Seismology and Earthquake Engineering (SEE7), IIEES, May 16-17, Tehran. Doi:

26. **Estekanchi H, Harati M, Mashayekhi M**.*, 2018, An Investigation on the interaction of RC shear walls and moment resisting frames in RC dual system using endurance time (ET) method, The Structural Design of Tall and Special Buildings, e1489, Wiley. Doi:* 10.1002/tal.1489
46

# AUTHOR BIOGRAPHIES

**Mojtaba Harati** is a lecturer at Civil Engineering Department of the University of Science and Culture, Rasht, Iran. He received his M.Sc. degree from SUT, Tehran, Iran. His research interests include performance-based earthquake engineering and structural topology optimization.

**Mohammadreza Mashayekhi** received his PhD degree from the Sharif University of Technology (SUT), Tehran, Iran. He is working as a postdoctoral research associate in the department of civil engineering at SUT. His research interests include structural engineering, earthquake engineering, continuum mechanics, and theory of elasticity.

**Ali Khansefid** is an assistance professor at K. N. Toosi University of Technology, Tehran, Iran, where he is currently working on two integrated projects, the first one is to propose a new model for generating random earthquake scenario which itself includes two separate model, the first one is a probabilistic model to generate earthquake events scenario in the certain period of time containing all probable main-shocks and aftershocks; and the second one is a model to generate a synthetic stochastic accelerograms for the Iranian plateau. In the second project, He is working on a new life-cycle based method for designing the active vibration control systems and evaluating their efficacy for risk mitigation in the probabilistic framework.

**Saeid Pourzeynali** is a professor of civil engineering at the University of Guilan, Rasht, Iran. He is a member of the Iranian Construction Engineers Organization, ASCE, Iranian Inventors Association, and several other professional associations. His research interests include a broad area of topics in structural and earthquake engineering with a special focus on the design of tall and bridge structures as well as topics that are more relevant to the structural control.

**Arash Bahar** is an associate professor of civil engineering at the University of Guilan, Rasht, Iran. His research interests include a broad area of topics in structural and earthquake engineering with a special focus on topics pertinent to the structural control.